\documentclass[%
 reprint,
 amsmath,amssymb,
prl,
 longbibliography,
 lengthcheck,%
]{revtex4-1}

\usepackage{graphicx}
\usepackage{dcolumn}
\usepackage{bm}
\usepackage{hyperref}

\makeatletter

\newcommand{\Rmnum}[1]{\expandafter\@slowromancap\romannumeral #1@}
\makeatother

\begin{document}

\preprint{APS/123-QED}

\title{Spin excitations in K$_{0.84}$Fe$_{1.99}$Se$_2$ superconductor as studied by M\"ossbauer spectroscopy}

\author{Zhiwei Li}
\author{Xiaoming Ma}
\author{Hua Pang}\email{hpang@lzu.edu.cn}
\author{Fashen Li}
 \affiliation{Institute of Applied Magnetics, Key Lab for Magnetism and Magnetic Materials of the Ministry of Education, Lanzhou University, Lanzhou 730000, Gansu, P.R. China.}

\date{\today}

\begin{abstract}
M\"ossbauer spectroscopy was used to probe the site specific information of the K$_{0.84}$Fe$_{1.99}$Se$_2$ superconductor. A spin excitation gap, $\Delta E \approx$5.5\,meV, is observed by analyzing the temperature dependence of the hyperfine magnetic field (HMF) at the iron site within the spin wave theory. Using a simple model suggested in the literature, the temperature dependence of the HMF is well reproduced, suggesting that, below room temperature, the alkali metal intercalated iron-selenide superconductors can be regarded as ferromagnetically coupled spin blocks that interact with each other antiferromagnetically to form the observed checkerboard-like magnetic structure.

\begin{description}
\item[PACS numbers]
76.80.+y, 74.10.+v
\end{description}
\end{abstract}

\maketitle


\renewcommand{\textbf}{\textrm}

\section{\label{sec:Intro}Introduction}
There has been a renewed interest in the iron-based superconductors since the discovery of superconductivity (SC) at about 30\,K in the $A_{0.8+\delta}$Fe$_{1.6+\beta}$Se$_{2}$ ($A$ = K, Rb, Cs or Tl/K) \cite{KFeSe-prb,RbFeSe32K,CsFeSe-MSC,TlFeSe-Mott} compounds due to their unprecedented physical properties, such as the coexistence of high temperature SC with strong antiferromagnetic (AFM) order \cite{CsFeSe-MSC,KMSC-neutron,CsMSC-neutron,Magn-MSC,Magnon-MSC}. However, whether SC and AFM order coexist microscopically or SC only occurs in the non-magnetic phase is still highly debated since some reports support the coexistence picture \cite{CsFeSe-MSC,KMSC-neutron,CsMSC-neutron,Magn-MSC,Magnon-MSC} and others favor the phase separation scenario \cite{PS-XRD,Ricci-XRD,PS-Moss,PS-ARPES}. Local probe techniques such as muon-spin relaxation/rotation ($\mu$SR) \cite{CsFeSe-MSC} and M\"ossbauer spectroscopy (MS) \cite{PS-Moss,Moss-Ryan,Moss-Nowik,Moss-Li} have shown that a two-component picture is inescapable to describe the system correctly, namely, all samples are phase-separated into a major AFM phase and a minor paramagnetic (PM) phase. Nuclear magnetic resonance (NMR) \cite{NMR-prl} and $\mu$SR \cite{uSR-PM} experiments reveal that the PM phase becomes superconducting below $T_c$. However, the scenario that only the PM phase becomes superconducting alone can not explain all the above mentioned experiments. Thus further studies on these compounds are desired to settle the debate. In this case, an investigation on the local magnetic property is helpful in understanding the correlation between SC and magnetic ordering of these systems.

MS has been proved to be a very useful tool to probe local specific information of the iron-based superconductors \cite{PS-Moss,Moss-Ryan,Moss-Nowik}. Especially, when possible coexistence of magnetic order and SC presents in the same sample, a MS study might reveal rich information. So far, only a few work using MS to study these materials \cite{PS-Moss,Moss-Ryan,Moss-Nowik} have been reported. A detailed study focusing on the temperature dependence of the local magnetic field at the iron site near the superconducting transition temperature is still missing, which might hold the key to understand the interesting interplay between AFM order and SC. Therefore, in the present work, MS was used to study the magnetic structure and temperature dependence of the hyperfine magnetic field (HMF) at the iron nucleus of K$_{0.84}$Fe$_{1.99}$Se$_2$ single crystals. The results provide evidence that a spin excitation gap opens up before entering the SC state. Using a simple spin model, we show that the ferromagnetically coupled (FMC) four spins can be viewed as a net spin, which couples antiferromagnetically with each other to form the checkerboard-like AFM structure.

\section{\label{sec:Experiment}Experiments}
Single crystals of potassium intercalated iron-selenides of nominal composition K$_{0.8}$Fe$_2$Se$_2$ were grown by the self melting method similar to previous reports \cite{KFeSe-prb,CsFeSe-MSC}. Stoichiometry of high purity K pieces, Fe and Se powders were mixed and put in a sealed quartz tube. The samples were heated to 1273\,K slowly, kept for 2\,h, cooled down to 973\,K at the rate of 5\,K/h and then furnace cooled to room temperature by shutting down the furnace. The resulting plate-like crystals with a shiny surface are of a size up to $6\times4\times2$\,mm$^3$. The actual composition is determined to be K$_{0.84}$Fe$_{1.99}$Se$_2$ by energy dispersive X-ray spectrum (EDXS).

Single crystal x-ray diffraction (XRD) measurements were performed on a Philips X'pert diffractometer with Cu K$_{\alpha}$ radiation. AC susceptibility measurements were carried out through a commercial (Quantum Design) superconducting quantum interference device (SQUID) magnetometer. Transmission M\"ossbauer spectra were recorded using a conventional constant acceleration spectrometer with a $\gamma$-ray source of 25\,mCi $^{57}$Co in palladium matrix moving at room temperature. The absorber was kept static in a temperature-controllable cryostat filled with helium gas. The velocity of the spectrometer was calibrated with $\alpha$-Fe at room temperature and all the isomer shift quoted in this work are relative to that of the $\alpha$-Fe.

\section{\label{sec:Results}Results and Discussion}
\begin{figure}[htp]
\includegraphics[width=8 cm]{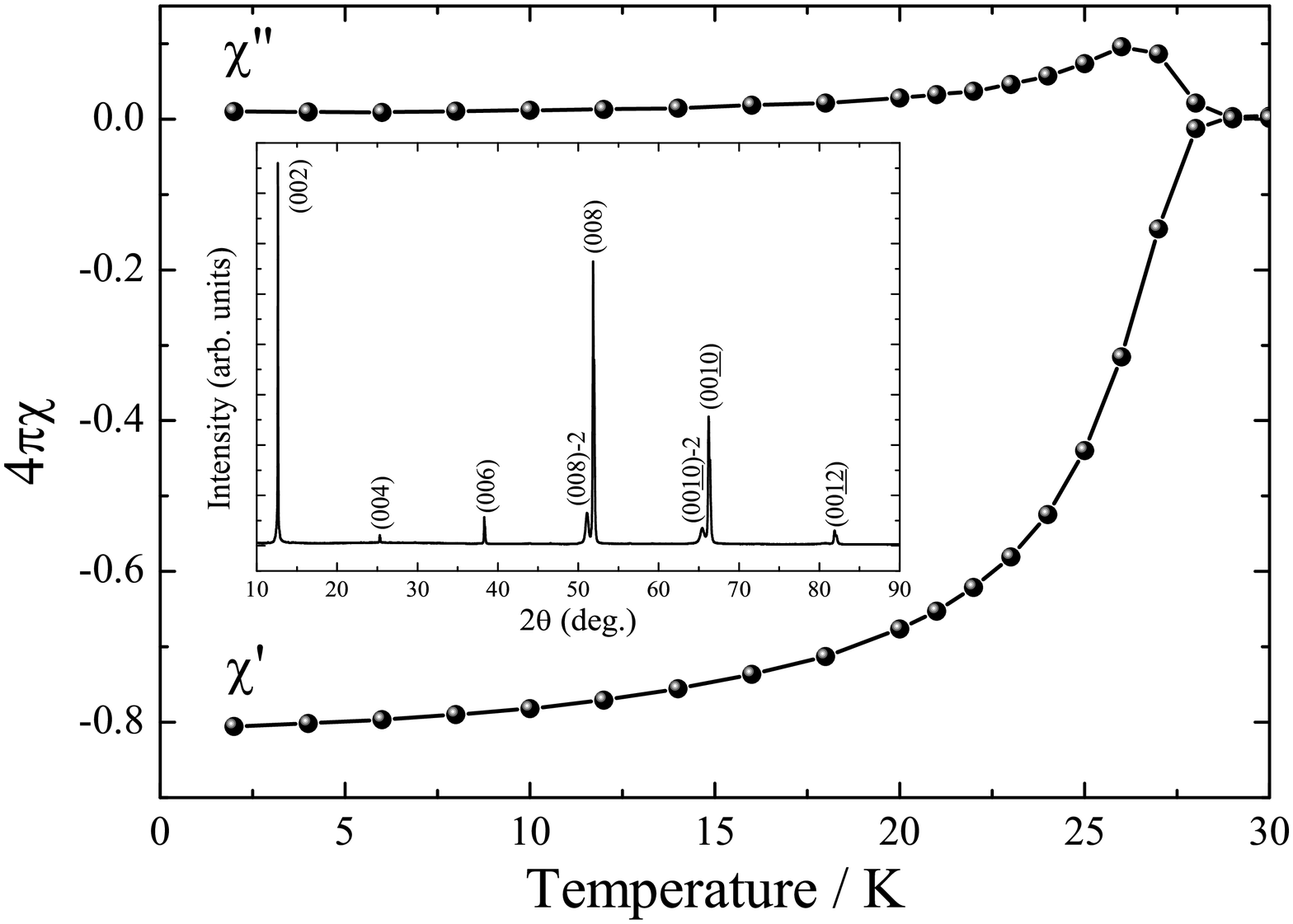}
\caption{\label{Fig1} Temperature dependence of AC susceptibility of the K$_{0.84}$Fe$_{1.99}$Se$_2$ crystal measured along the $ab$-plane with $H_{ac}$=1\,Oe and $f$=300\,Hz. Inset is the single crystal X-ray diffraction pattern of K$_{0.84}$Fe$_{1.99}$Se$_2$ crystal.}
\end{figure}

Single crystal x-ray diffraction pattern of K$_{0.84}$Fe$_{1.99}$Se$_2$ is shown in the inset of Fig. \ref{Fig1}. As can be seen, only ($00l$) diffraction peaks are observed, indicating the crystallographic $c$-axis is perpendicular to the plane of the plate-like single crystal. Interestingly, two sets of ($00l$) reflections corresponding $c_1$=14.098\,{\AA} and $c_2$=14.272\,{\AA} are observed, which are attributed to the inhomogeneous distribution of the intercalated K atoms \cite{AFeSe-TwoSet}. The temperature dependence of the AC susceptibility of K$_{0.84}$Fe$_{1.99}$Se$_2$ single crystal measured along the $ab$-plane with $H_{ac}$=1\,Oe and $f$=300\,Hz is shown in Fig. \ref{Fig1}. The onset superconducting transition temperature, $T_c$, is determined to be 28\,K from the real part of the susceptibility. The superconducting volume fraction is estimated to be $\sim$80\% at 2\,K.

\begin{figure}[htp]
\includegraphics[width=8 cm]{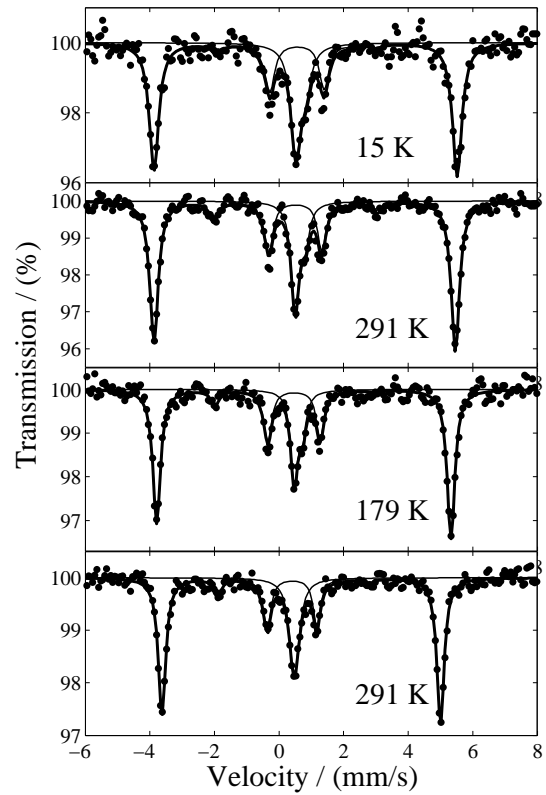}
\caption{\label{Moss} M\"ossbauer spectra taken at indicated temperatures of the K$_{0.84}$Fe$_{1.99}$Se$_2$ single crystal. The M\"ossbauer absorber was prepared with well-cleaved flake-like single crystals, which were put together with the $c$-axis aligned perpendicular to the plane of the M\"ossbauer absorber.}
\end{figure}

M\"ossbauer spectra of a mosaic of single crystal flakes, oriented on a thin paper underlayer so that the $c$-axis is perpendicular to the plane of the M\"ossbauer absorber, recorded below room temperatures are shown in Fig. \ref{Moss}. All spectra share similar spectral shapes and are fitted with two components: a dominant magnetic sextet and a nonmagnetic quadrupole doublet, with \textsc{MossWinn 4.0} \cite{MossWinn} programe.

Make an intense study of the doublet, one finds that the two peaks are strongly polarized with an intensity ratio close to 3:1. This means that the orientation of the main axis of the electric field gradient (EFG) of the PM phase is parallel to the $c$ axis of the crystal, which agrees well with the fitted angle of $\theta_{pm}$=8(3)$^{\circ}$ by V. Ksenofontov \textit{et al} \cite{PS-Moss}. Due to the low statistical of the data and to simplify the fitting procedure, we fitted the doublet assuming that $\theta_{pm}$=0$^{\circ}$. The derived isomer shift and quadrupole splitting at 15\,K is found to be $\delta$=0.631\,mm/s and $eQV_{zz}/2$=-0.272\,mm/s, respectively. These hyperfine parameters are close to the reported values of $\beta$-FeSe \cite{FeSe-PseudoPhase,MossFeSe}, while the quadrupole splitting is a little bit smaller than the corresponding doublet in the Rb$_{0.8}$Fe$_{1.6}$Se$_2$ compound \cite{PS-Moss}, which might be due to different stoichiometries of these samples. Therefore, we may attribute the doublet to the FeSe phase (pseudo-FeSe phase), which corresponds to the FeSe$_4$ tetrahedrons that have K vacancy neighbors in the crystal structure. The coexistence of nonmagnetic pseudo-FeSe phase with the main AFM phase can be naturally understood in the phase separation scenario, which is supported by scanning nanofocused x-ray diffraction \cite{PS-XRD,Ricci-XRD}, NMR \cite{NMR-prl}, and previous M\"ossbauer \cite{PS-Moss} studies.

In order to get a better fit of the spectra, special care should be taken in adjusting the sextet. In our previous manuscript \cite{Moss-Li}, we fitted the spectra with a relatively small EFG value, assuming that the axis of the main component of the EFG coincide with the crystallographic $c$-axis and the direction of the magnetic moments of the Fe atoms. A close inspection of the spectra reveals that this procedure can not account for the slightly asymmetries of the line pairs (1,6) and (3,4) and the positions of the line pairs (2,5) and (3,4) as pointed out by V. Ksenofontov \textit{et al} \cite{PS-Moss}. Therefore, in the present paper we refitted the spectra according to the procedures given by V. Ksenofontov \textit{et al}, which solves the static Hamiltonian for mixed magnetic and quadrupole interactions with arbitrary relative orientation. The asymmetry parameter $\eta=(V_{xx}-V_{yy})/V_{zz}$ is assumed to be zero to further simplify the problem since the fitted value for Rb$_{0.8}$Fe$_{1.6}$Se$_2$ is rather small $\sim0.1$ \cite{PS-Moss}. Fitting the spectra yields an averaged relative intensity of 75\% for the AFM phase, which is significantly smaller than the reported value of 88\% for Rb$_{0.8}$Fe$_{1.6}$Se$_2$ \cite{PS-Moss} and K$_{0.8}$Fe$_{1.76}$Se$_2$ \cite{Moss-Ryan} compounds. This may be caused by the high amounts of iron in our K$_{0.84}$Fe$_{1.99}$Se$_2$ crystal, which may favor the pseudo-FeSe phase. The derived hyperfine parameters for the AFM phase are $\delta$=0.654\,mm/s and $eQV_{zz}/2$=1.172\,mm/s with an angle $\theta_{afm}$=44(1)$^{\circ}$ between the axis of $V_{zz}$ and the HMF $B_{hf}$=28.32\,T at 15\,K, compares well with previously reported values of some similar compounds \cite{PS-Moss,Moss-Nowik}.

\begin{figure}[htp]
\includegraphics[width=8 cm]{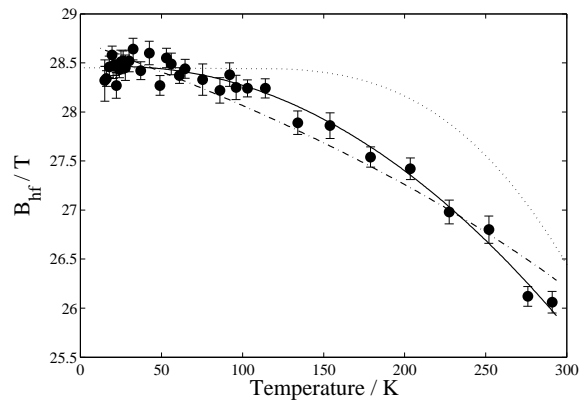}
\caption{\label{HF} Temperature dependence of the hyperfine field, $M(T)$, extracted from least-squares fits of the M\"ossbauer spectra. Fitting to the $M(T)$ data with different theories are also compared, the power law (dashdotted line), Brillouin function (dot line) and gaped spin wave theory (solid line). As can be seen, the 3D spin wave theory with an energy gap of $\Delta E \sim$5.5\,meV can better reproduce the low-temperature plateau of the hyperfine field (see text).}
\end{figure}

In order to get a better understanding of the magnetic properties in these materials, we investigated the temperature dependence of the AFM order parameter. The temperature dependence of the HMF, $B_{hf}(T)$, at the Fe site in K$_{0.84}$Fe$_{1.99}$Se$_2$ is depicted in Fig. \ref{HF}, together with different fitting results. Similar to the behavior of neutron powder diffraction (NPD) (101) magnetic Bragg peak intensity profile \cite{KMSC-neutron}, the HMF shows an plateau below $\sim50\,K$ and then decreases gradually with increasing temperature. A simple Brillouin function was used to fit the HMF data in a previous work \cite{Moss-Ryan} and a rough agreement was found in the temperature range 10-530\,K. However, as can be seen from Fig. \ref{HF}, the Brillouin function (dot line) together with the power law (dashdotted line) fails to describe the low temperature behavior of the HMF. As is well known, in the temperature range of $T\ll T_N$, the decrease in HMF with increasing temperature can be well explained by spin excitations \cite{SpinWave-book} within the spin wave theory. For a three dimensional antiferromagnet, the temperature dependence of HMF at low temperatures follows \cite{SpinWave-book2},
\begin{eqnarray}
B_{hf}(T) = B_{hf}(0)(1 - C T^{2}e^{-\Delta E/k_BT})
\label{MSEG}
\end{eqnarray}
where $C$ is a constant that contains the spin wave stiffness. $\Delta E$ is the spin excitation gap (SEG), which is necessary to reproduce the plateau of the HMF at low temperatures. Applying equation (\ref{MSEG}) to the data yields the following results, $B_{hf}(0)$=28.47\,T, $\Delta E$=63\,K ($\sim$5.5\,meV). Interestingly, the fitted $\Delta E\sim$5.5\,meV has a nonzero value, suggesting that a substantial SEG due to spin anisotropy opens up above the the superconducting transition temperature. Actually, the SEG has been predicted theoretically \cite{SpinWave-Theory} and was recently observed by neutron scattering studies in Rb$_{0.89}$Fe$_{1.58}$Se$_2$ \cite{SEG-RbFeSe} compound.  SEG was also observed in YBa$_2$Cu$_3$O$_{7-\delta}$ (YBCO) and La$_{2-x}$Sr$_x$CuO$_4$ (LSCO) \cite{SEG-Cuprates,Kofu-SEG,Millis-SEG1993,Millis-SEG1994,Bourges-SEG,Anderson-SEG} cuprate superconductors and whether the SEG is related to superconductivity is still an open question. There is experimental evidence that well-defined SEG ($\sim$6\,meV) in the incommensurate spin fluctuations is observed in the superconducting state only for samples close to the optimal doping \cite{optimaldoping} for LSCO systems. And a rough proportionality between T$_C$ and SEG was also observed for YBCO superconductors: in the weakly doped region, $E_G\approx k_BT_C$, while in the heavily doped region, $E_G\approx 3.8k_BT_C$ \cite{SEGPropTC,Dai2001,Regnault1994}. Thus, SEG and SC might closely related with each other and a thorough investigation of the evolution of SEG and $T_C$ with different carrier-doping levels is highly desired. In this aspect, in-depth neutron scattering studies of the AFM spin excitation spectrum may yield fruitful information on this issue.

To understand the temperature dependence of the HMF and estimate the magnetic exchange interactions of our sample, we go to the novel magnetic ordering structure of this compound below $T_N$. The magnetic moments of the four irons in each $\sqrt{5}\times\sqrt{5}$ unit cell align ferromagnetically along the crystalline $c$-axis \cite{KMSC-neutron,PS-Moss}. And the ferromagnetic blocks interact with each other antiferromagnetically to form a block-checkerboard AFM pattern. Though accredited values of exchange interactions in this system are not reached, strong interactions within the FMC blocks have been predicted theoretically and observed experimentally \cite{SpinWave-Theory,SEG-RbFeSe}. A recent neutron scattering experiment showed that the acoustic spin waves between $\sim$9\,meV to $\sim$70\,meV arise mostly from AFM interactions between the FMC blocks, while the optical spin waves associated with the exchange interactions of iron spins within the FM blocks are above $\sim$80\,meV \cite{SEG-RbFeSe}.

\begin{figure}[htp]
\includegraphics[width=8 cm]{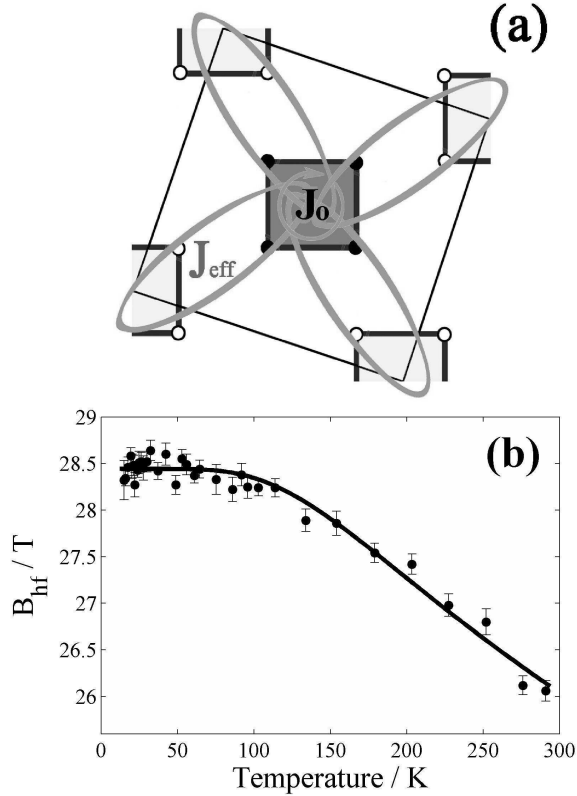}
\caption{\label{Heisenberg} (a) schematic representation of the effective interaction model used in the text. $J_0$ represents the effective interaction constant within each FMC block and is strong enough with respect to the inter-block interaction constant $J_{eff}$ to force the four spins fluctuate coherently at finite temperatures. (b) temperature dependence of the HMF together with fitting results using equation (\ref{BhfT}) (see text).}
\end{figure}

Considering the energy scales of the acoustic and optical spin waves and the large energy separation between them, it is reasonable to assume that, at low temperatures, the decrease in HMF with increasing temperature below room temperature is controlled by the AFM interactions of the FMC blocks. Thus, by fitting the temperature dependence of the HMF data we can deduce the effective interaction, $J_{eff}$, between two nearest FMC blocks, as illustrated in Fig. \ref{Heisenberg} (a). To simplify the calculation, we make a premise that the iron spins in the FMC block fluctuate coherently. This is reasonable at least at temperatures below room temperature due to the strong ferromagnetic interactions within the FMC block. In this case, an FMC block can be regarded as a super-spin with $S_{eff}=8$.

In the simplest case of two interacting spins, the energy levels are $E(S)=J_{eff}S(S+1)$, where $\vec{S}=\vec{S_1}+\vec{S_2}$ and $\vec{S_1}$ and $\vec{S_2}$ are the angular momenta of the two coupled spins. The HMF probed by M\"ossbauer measurements will be $B_{hf}=C\langle S_Z\rangle$, where $\langle S_Z\rangle$ is the expectation value of the $z$ component of $\vec{S_i}$, and reaches its maximum value at ground state (zero temperature). While at elevated temperatures, states with higher $S$ are accessible, which results in the decrease of HMF as observed above. If we assume the relaxation between the electronic states is fast with respect to the Larmor precession time, we can express the finite temperature HMF as $B_{hf}(T)=B_{hf}(0)(1-\sum_Sh_Sn_S)$, and
\begin{eqnarray}
\small
\sum_Sh_Sn_S= \frac{\sum\limits_{S=0}^{16}\sum\limits_{S_z=0}^S h(S,S_Z)e^{-J_{eff}S(S+1)/k_BT}}{\sum\limits_{S=0}^{16}\sum\limits_{S_z=-S}^S e^{-J_{eff}S(S+1)/k_BT}},
\label{BhfT}
\end{eqnarray}
where $n_S$ is the populations and $h(S,S_Z)$ is the decrease in HMF corresponding to state $|S,S_Z\rangle$. If we assume that $h(S,S_Z)$ is proportional to $S_Z$ with the same proportionate constant for all $S$ states, $h(S,S_Z)=h_0S_Z$, then we can fit the experimental data with equation (\ref{BhfT}). As can be seen from Fig. \ref{Heisenberg} (b), a good agreement between the theoretical curve and the experimental data can be obtained and the fitted parameters are $J_{eff}$=22.8\,meV, $B_{hf}(0)$=28.44\,T and $h_0$=0.697\,T.

To see the efficiency of our simple model in describing the low-energy spin excitations, we compare our results with that deduced from the effective spin Hamiltonian model, which has been widely used to describe the ground state and spin excitations for this type of compounds. Usually, the Hamiltonian involves intra-block nearest and second nearest neighbor interactions $J_1$, $J_2$ and the inter-block nearest and second nearest neighbor interactions $J_1'$, $J_2'$. Even the third nearest neighbor interactions $J_3$, $J_3'$ have been adopted to fit the spin wave spectra by Miaoyin Wang \textit{et al} \cite{SEG-RbFeSe}. In terms of the $J_1$-$J_1'$-$J_2$-$J_2'$-$J_3$-$J_3'$ model, the low-energy spin waves can be approximately described by ($J_1'+2J_2'+2J_3)S/4\sim$17\,meV. Obviously, our results of $J_{eff}$ agrees reasonably well with the neutrons scattering results, which proves the validity of our above assumption.

\section{\label{sec:Conclusion}Concluding remarks}
High quality single crystals of K$_{0.84}$Fe$_{1.99}$Se$_2$ have been prepared and studied by M\"ossbauer spectroscopy. Temperature dependence of the hyperfine magnetic field is well explained within the gaped spin wave theory. Fitting the experimental data yields a spin excitation gap of about 5.5\,meV/63\,K. Supposing the blocked spins fluctuate coherently, the effective exchange interaction between these coupled spin blocks is estimated to be $J_{eff}$=22.8\,meV, which agrees reasonably well with previous NPD estimated value ($J_{eff}\sim$17\,meV).

\begin{acknowledgments}
We thank T. Xiang, P. C. Dai and Y. Z. You for useful discussions.
This work was supported by the National Natural Science Foundation of China under Grants No. 10975066.

\end{acknowledgments}


\end{document}